# Bifurcations of two coupled classical spin oscillators

B.Rumpf and H.Sauermann
Technische Hochschule Darmstadt
Institut für Festkörperphysik

**Abstract**

Two classical, damped and driven spin oscillators with an isotropic exchange interaction are considered. They represent a nontrivial physical system whose equations of motion are shown to allow for an analytic treatment of local codimension 1 and 2 bifurcations. In addition, numerical results are presented which exhibit a Feigenbaum route to chaos.

# Introduction

Within the last few years several studies concerning the dynamical behaviour of classical hamiltonian spin systems have been published. The motion of a single spin being exposed to an external driving field as well as an anisotropy field was discussed in [1]. Questions of integrability for coupled spin systems under the influence of single site as well as exchange anisotropy were thoroughly explored in [2] [3] and also in [4].
In many cases dissipative effects cannot be neglected, however. In order to investigate their influence we study as a paradigmatic example the motion of two coupled, damped spin oscillators in a static field $B_z$ being driven by a rotating field $B_0$ in the x-y-plane. In agreement with microscopic derivations [5] as well as phenomenological treatments [6] the damping is described by adding a Gilbert-term to the Landau-Lifshitz- equations of motion. Such a situation may also be interpreted in terms of two interacting homogenous modes in a magnetic sample. Despite its physical simplicity - it is in fact the simplest nontrivial situation if the driving field is circularly polarized and there are no anisotropy fields - it will be shown to exhibit a rather complex bifurcational behaviour. A similar system which however doesn't possess permutation symmetry has been studied in [7] with the main purpose of deriving an effective g-factor.

# Bifurcation analysis

For simplicity and to clarify the notation we start with the trivial case of the motion of a single spin

$$-\dot{\vec{S}} = \vec{S} \times \vec{H} + \Gamma \vec{S} \times (\vec{S} \times \vec{H}) \tag{1}$$

where

$$\vec{H} = B_0(\vec{e}_x \cos\omega t + \vec{e}_y \sin\omega t) + B_z \vec{e}_z \tag{2}$$

and $\Gamma$ is the damping parameter. Noting that equation (1) conserves modulus, we introduce a rotating coordinate system $\vec{e}_x' = \vec{e}_x \cos\omega t + \vec{e}_y \sin\omega t, \vec{e}_y' = -\vec{e}_x \sin\omega t + \vec{e}_y \cos\omega t, \vec{e}_z' = \vec{e}_z$, drop the prims after the transformation and project the sphere $\mid \vec{S} \mid = 1$ into a complex plane. This method is well known in the present context, see e.g.[8],[9]. We find the following differential equation for the complex variable $z = \frac{S_x + iS_y}{1+S_z}$ :

$$\dot{z} = (i - \Gamma)B_z z - i\omega z - (i - \Gamma)\frac{B_0}{2}(1 - z^2) \tag{3}$$

This equation yields a stable and an unstable fixed point $z_{A/B} = -\frac{(i-\Gamma)B_z - i\omega}{(i-\Gamma)B_0} \pm \sqrt{(\frac{(i-\Gamma)B_z - i\omega}{(i-\Gamma)B_0})^2 + 1}$ with the eigenvalues $\xi_{1,A/B} = (i - \Gamma)(B_z + B_0 z_{A/B}) - i\omega, \xi_{2,A/B} = \bar{\xi}_{1,A/B}$ of the corresponding Jacobi matrices. As to the projection technique the reader is refered to the appendix where the most general version in our context is given. Our basic Landau-Lifshitz-equation for two driven, isotropically coupled classical spins are given by

$$-\dot{\vec{S}}_{1/2} = \vec{S}_{1/2} \times \vec{H}_{eff1/2} + \Gamma \vec{S}_{1/2} \times \vec{S}_{1/2} \times \vec{H}_{eff1/2} \tag{4}$$

where the effective fields

$$\vec{H}_{eff1/2} = B_0(\vec{e}_x \cos\omega t + \vec{e}_y \sin\omega t) + B_z \vec{e}_z + J\vec{S}_{2/1} \tag{5}$$

now contain the interaction terms $J\vec{S}_{1/2}$. These equations are invariant under permutation of the two spin indices. Transformation into a rotating coordinate system and stereographic projection as before yields

$$\begin{aligned} \dot{z}_1 &= (i-\Gamma)B_z z_1 - i\omega z_1 - (i-\Gamma)\frac{B_0}{2}(1-z_1^2) + J(i-\Gamma)(z_1-z_2)\frac{1+z_1\bar{z}_2}{1+z_2\bar{z}_2} \\ \dot{z}_2 &= (i-\Gamma)B_z z_2 - i\omega z_2 - (i-\Gamma)\frac{B_0}{2}(1-z_2^2) + J(i-\Gamma)(z_2-z_1)\frac{1+z_2\bar{z}_1}{1+z_1\bar{z}_1} \end{aligned} \tag{6}$$

This four dimensional system of equations is invariant under time reversal together with a simultaneous change of $(B_z, B_0, \omega, J, \Gamma)$ to $(-B_z, -B_0, -\omega, -J, \Gamma)$. So, having obtained e.g. a solution for positive $J$ (ferromagnetic coupling) one gets a corresponding solution with opposite stability for negative $J$ (antiferromagnetic coupling) just by considering a time reversed situation with opposite directions of the magnetic fields. Note that one of the five parameters may be eliminated by introducing dimensionless quantities.
It follows immediately from equations (6), that the spins do not interact, if they are parallel to each other. So there are two trivial fixed points

$$z_1 = z_2 = z_A \tag{7}$$

$$z_1 = z_2 = z_B \tag{8}$$

in four dimensional phase space. The eigenvalues of the corresponding Jacobi matrices are

$$\lambda_{1/2} = \mathrm{Re}\xi \pm i\mathrm{Im}\xi \tag{9}$$

$$\lambda_{3/4} = \mathrm{Re}\xi - 2J\Gamma \pm i(\mathrm{Im}\xi + 2J) \tag{10}$$

where $\xi = \xi_A$ or $\xi_B$ with $\xi_{A/B} = (i-\Gamma)(B_z + B_0 z_{A/B}) - i\omega$. Note that $\xi_A$ and $\xi_B$ are the eigenvalues of the single spin calculated above. At this stage one proves, that for $\mathrm{Re}\xi - 2J\Gamma = 0$ and $\mathrm{Im}\xi + 2J \neq 0$ a Hopf bifurcation occurs. The resulting limit cycle is stable (unstable) for $J < 0 (J > 0)$. In both cases a center manifold reduction of equations (6) which has been peformed in the settings of ref.s [10],[11] respectively[1] yields for the Hopf normal form

$$\dot{z} = \lambda_3 z + \gamma z^2 \bar{z} \tag{11}$$

where $z = z_1 - z_2$ parametrizes the 2 dimensional center manifold and $\gamma = -\frac{(i-\Gamma)J}{(1+z_{A/B}\bar{z}_{A/B})^2}$. As usual the unfolding parameter is the real part of $\lambda_3$. The corresponding bifurcation lines $H_1$ and $H_2$ are plotted in the $\omega$-J-plane in figures (1) and (2).

For $B_0^2 = 4J^2 + \Gamma^2 B_z^2$ and $\omega = (1+\Gamma^2)B_z$ the eigenvalues $\lambda_{3/4}$ vanish altogether. Performing again a center manifold reduction the following differential equation is obtained

$$\dot{z} = \lambda_3 z + \delta z^3 + \gamma z^2 \bar{z} \tag{12}$$

where
$\gamma = -\frac{J}{4}(i-\Gamma)$ and $\delta = -J(i-\Gamma)\frac{\Gamma^2 B_z^2}{4B_0^2}$. We have a codimension 2 singularity which is unfolded by the 2 dimensional complex parameter $\lambda_3$. Further analysis of this normal form shows that it describes a situation in which two saddle node bifurcation lines of nontrivial fixed points given by $\mid \delta\lambda_3 \mid - \mid \mathrm{Im}(\lambda_3\bar{\gamma}) \mid = 0$ , $\mathrm{Re}\lambda_3 > 0$ (for $J < 0$) meet the Hopf bifurcation line (see point A in figures (1) and (2)).

We now turn to the nontrivial fixed points for which $\vec{S}_{fix1} \neq \vec{S}_{fix2}$.For vanishing coupling strength there are only two such fixed points being related to one another by permutation of the two spins. We shall see that there may be as many as four more fixed points for finite coupling. In order to prove this assertion we develop an analytical method which permits not only to solve the fixed point problem of equations (6) but also to discuss the pertaining question of stability. We introduce a new coordinate system $\vec{e}_\xi, \vec{e}_\eta, \vec{e}_\zeta$ and project stereographically along the $\vec{e}_\zeta$-axis into its $\vec{e}_\xi - \vec{e}_\eta$-plane. The initially unknown directions of these unit vectors are determined unambiguously by requiring that $\vec{e}_\zeta$ and $\vec{e}_\xi$ are proportional to $\vec{S}_1 + \vec{S}_2$ and $\vec{S}_1 - \vec{S}_2$ respectively. As a consequence the fixed points are located symmetrically with respect to the origin of the $\vec{e}_\xi - \vec{e}_\eta$-plane at the points $\pm x$ of the $\vec{e}_\xi$-axis. We prove in the appendix that the square of the total magnetisation $u = \frac{1}{4}(\vec{S}_1 + \vec{S}_2)^2 = (\frac{1-x^2}{1+x^2})^2$, which means that $0 \leq u \leq 1$, is obtained by solving the following third order equation

$$\Gamma^2\omega^2(1+u)^2(B_0^2 - 4J^2 u) = [(1+\Gamma^2)B_0^2 + (B_z - \omega)^2 + \Gamma^2 B_z^2 - 4J^2(1+\Gamma^2)u]^2 u \tag{13}$$

Having found a solution of this equation, each of which corresponds to two fixed points for the spins being connected by permutation symmetry, its orientation i.e. that of $\vec{e}_\xi, \vec{e}_\eta$ and $\vec{e}_\zeta$, may be calculated with the help of the the formulas given in the appendix.

[1]Incidentally as far as bifurcation theory is concerned we restrict ourselves by just referring to these two books. A useful and more recent overview giving a large though nevertheless selected list of references may be found in [12].

In order to investigate possible bifurcations connected with these nontrivial fixed points as well as their stability we derive first of all an explicit expression for the corresponding ( $4 \times 4$ )-Jacobi matrices. Working in the new coordinate system, putting $z_{1/2} = \pm x + \Delta z_{1/2}$ and considering equations (21)-(24) of the appendix one has the four equations

$$\begin{aligned}
\Delta \dot{z}_{1/2} &= \pm 2(a+bi)x\Delta z_{1/2} + (a+bi)\Delta z_{1/2}^2 - 2J(i-\Gamma)\frac{1-x^2}{1+x^2}\Delta z_{1/2} \\
&\pm J(i-\Gamma)[(2x+\Delta z_1 - \Delta z_2)\frac{1+(\pm x+\Delta z_{1/2})(\mp x+\Delta \bar{z}_{2/1})}{1+(\mp x+\Delta z_{2/1})(\mp x+\Delta \bar{z}_{2/1})} \\
&-2x\frac{1-x^2}{1+x^2}] \qquad (14)
\end{aligned}$$

Now any further calculations would become very tedious if one continued to work with $B_0, B_z$ and $\omega$ as independent parameters. This is firstly due to the fact that $a, b$ and $x$ are rather complicated functions of the old parameters and secondly that one would have to discuss different Jacobians for different fixed points. Therefore in accordance with the concluding remarks of the appendix we consider from now on $a, b$ and $x$ as independent bifurcation parameters (leaving J and $\Gamma$ unchanged) as far as analytical considerations are concerned. However, in what follows we switch back to the old parameters for physical reasons when displaying our results graphically. Linearisation of equation (14) yields

$$\begin{pmatrix}
2ax+J\Gamma & -2bx+J & -J\Gamma q & J \\
2bx-J & 2ax+J\Gamma & Jq & J\Gamma \\
-J\Gamma q & J & -2ax+J\Gamma & 2bx+J \\
Jq & J\Gamma & -2bx-J & -2ax+J\Gamma
\end{pmatrix} \qquad (15)$$

for the Jacobian. The corresponding eigenvalue equation reads

$$\lambda^4 + K_3\lambda^3 + K_2\lambda^2 + K_1\lambda + K_0 = 0$$

with

$$\begin{aligned}
K_3 &= -4\Gamma J \\
K_2 &= 8x^2(b^2-a^2) + J^2(2-2q+5\Gamma^2-\Gamma^2q^2) \\
K_1 &= 32x^2abJ + 2J\Gamma[8x^2(a^2-b^2) + (1+\Gamma^2)(q^2-1)J^2] \\
K_0 &= 16x^4(a^2+b^2)^2 + J^2x^2[8a^2(1+q) + 4b^2(q^2-1) - 24ab\Gamma + 4a^2\Gamma^2(q^2-1) + \\
&8b^2\Gamma^2(q+1) + 8ab\Gamma(q^2-2q)]
\end{aligned} \qquad (16)$$

where $q = \frac{8x^2}{(1+x^2)^2} - 1$, i.e. $-1 \leq q \leq 1$. Using these formulas it is now possible to write down the explicit algebraic expressions for the conditions which have to be met for the various local bifurcations to occur. They depend on altogether 5 parameters $a, b, x, J, \Gamma$ and read for saddle-node and Hopf-bifurcations $K_0 = 0$ and $K_1^2 - K_1K_2K_3 + K_0K_3^2 = 0, K_1K_3 > 0$ respectively. The corresponding bifurcation lines together with the Hopf line of the trivial fixed point are plotted in an $\omega - J$-plane in figures (1) and (2) by evaluating the algebraic expression with a computer. They are symmetrical to the axis $J = 0$. We find two cusp points $C_1$ and $C_2$. We recognize point A as the codimension 2 point which was already discussed in connection with equation (12). There are six nontrivial fixed points inside the triangle A,$C_1, C_2$ and two outside respectively.

Next we point out that all of the three basic cases of local codimension 2 bifurcations of ref.[10], chapter 7, occur in our system. There is a double zero eigenvalue associated with linear part

$$\begin{pmatrix} 0 & 1 \\ 0 & 0 \end{pmatrix} \tag{17}$$

at $F_1$ and $F_2$, where saddle node and Hopf lines meet. One finds a pure imaginary pair and a simple zero with linear part

$$\begin{pmatrix} 0 & 0 & 0 \\ 0 & 0 & -\omega \\ 0 & \omega & 0 \end{pmatrix} \tag{18}$$

at the points D in fig.(2). The situation of fig.(2) is obtained from that of fig.(1) by decreasing the field $B_0$. Thereby $F_1$ approaches A and reaches it at a certain value of $B_0$ and coincides with A later on. Before that happens the two points $D_1$ and $D_2$, in which the Hopf and saddle-node lines meet tangentially, appear simultaneously, they persist thereafter.

An analytic argument for the existence of these types of bifurcation is the following. A double zero eigenvalue exists if and only if $K_0 = K_1 = 0$. This is true for $a = 0, 4b^2x^2 + J^2(q^2 - 1) = 0$ at $\Gamma = 0$. Because

$$\begin{vmatrix} \frac{\partial K_0}{\partial a} & \frac{\partial K_0}{\partial b} \\ \frac{\partial K_1}{\partial a} & \frac{\partial K_1}{\partial b} \end{vmatrix} \neq 0, \tag{19}$$

the implicit function theorem tells us, that there are (for $x$ and $J$ fixed) unique solutions $a(\Gamma)$ and $b(\Gamma)$ in the neighbourhood of that point. For the second bifurcation in question the two conditions have to be replaced by $K_0 = K_3K_2 - K_1 = 0, K_2 > 0$, the remainder goes through unchanged. In a three dimensional parameter space the codimension 2 bifurcations are represented by lines instead of points. They are plotted in fig.(3). We draw the attention of the reader to the points $V_1$ and $V_2$ where $F_1$ and $F_2$ meet line A. At present we desist from discussing them any further.

We finally state that there are two imaginary pairs of eigenvalues and a corresponding four dimensional center manifold at point E in figure (1),(2).

## Chaotic dynamics

We do not investigate any further the global bifurcations which usually occur in connection with the local codimension 2 bifurcations of the last section. Nevertheless it may be expected that they are an important cause for chaotic behaviour in our system. What we have done however in this respect is to employ the software-package AUTO [13] that can follow unstable limit cycles to get at least some numerical survey. We found that the stable limit cycles grown out from the nontrivial fixed points for $J < 0$ by Hopf bifurcations pass through a period doubling scenario which ends up in a chaotic attractor (fig.(4)). The Lyapunov spectrum as a function of the coupling parameter is plotted in fig.(5). The two Lyapunov exponents which are not shown are strictly negative. As is characteristic for such cases of broken symmetry [16] there are two homologous Feigenbaum routes related to each other by permuting the two spins. For $J > 0$ there is a corresponding repellor. Increasing $\mid J \mid$ further we get - after a sequence of periodic windows - a reverse Feigenbaum route. It leads after a final pitchfork bifurcation to a stable cycle which then persists forever. It is remarkable that the symmetric limit cycle which has been created originally, i.e. for small $\mid J \mid$ by a Hopf-bifurcation of the trivial fixed point and has changed its stability character in a subsequent global bifurcation, is the necessary unstable object in that last bifurcation. This was demonstrated independently by observing the continuous deformation of the final cycle into the initial one along a completely different path in parameter space.

## Summary

We have proven that the four dimensional equations of motion of two coupled damped spin oscillators which are driven by a rotating field lead to a surprising variety of dynamical and bifurcational phenomena. One finds all the basic codimension 1 and 2 bifurcations as well as a route to chaos. As our system represents a model for an antiferromagnet with constant sublattice coupling it provides an important physical example for a case study of this sort of problems. Our results differ strongly from the trivial ones for a single spin oscillator as well as those for two coupled but undamped spins [2],[3]. The same is true if we compare them with those derived in [17] for two torsionally coupled damped and driven pendula.

## Acknowledgement

The authors would like to thank Mr.T.Träxler for valuable discussions and comments. This work was performed within a program of the Sonderforschungsbereich 185 Nicht-lineare Dynamik, Darmstadt, Frankfurt. One of us (B.R.) was supported by Graduiertenförderung des Landes Hessen.

# Appendix

Projecting the following equation

$$-\dot{\vec{S}}_1 = \vec{S}_1 \times (\vec{B} - \vec{\omega} + J\vec{S}_2) + \Gamma \vec{S}_1 \times (\vec{S}_1 \times (\vec{B} + J\vec{S}_2)) \tag{20}$$

along the arbitrary direction $\vec{e}_\zeta = \vec{e}_\xi \times \vec{e}_\eta$ into the $\vec{e}_\xi - \vec{e}_\eta$ plane yields

$$\begin{aligned} \dot{z}_1 &= (i-\Gamma)B_\xi \frac{z_1^2-1}{2} - i(i-\Gamma)B_\eta \frac{1+z_1^2}{2} + (i-\Gamma)B_\zeta z_1 \\ &\quad -i\omega_\xi \frac{z_1^2-1}{2} - \omega_\eta \frac{1+z_1^2}{2} - i\omega_\zeta z_1 \\ &\quad +J(i-\Gamma)(z_1-z_2)\frac{1+z_1\bar{z}_2}{1+z_2\bar{z}_2} \end{aligned} \tag{21}$$

$B_\xi, B_\eta, B_\zeta$ and $\omega_\xi, \omega_\eta, \omega_\zeta$ are the new components of $\vec{B} = B_0\vec{e}_x + B_z\vec{e}_z$ and $\vec{\omega} = \omega\vec{e}_z$ in the directions $\vec{e}_\xi, \vec{e}_\eta$ and $\vec{e}_\zeta$ respectively. Choosing the axes $\vec{e}_\xi, \vec{e}_\eta, \vec{e}_\zeta$ as described in the text the resulting fixed point equations take the form

$$0 = Ax^2 \pm Bx + C \pm J(i-\Gamma)\frac{1-x^2}{1+x^2} \tag{22}$$

Explicit expressions for the complex quantities A,B,C follow by comparison with (21). Adding and subtracting these two equations and dropping the trivial solution $x = 0$ leads to

$$0 = Ax^2 + C \tag{23}$$

$$0 = (B + 2J(i-\Gamma)\frac{1-x^2}{1+x^2}) \tag{24}$$

Obviously equation (24) requires B to be equal to $(i-\Gamma)$ times some real number which turns out to be $B_\zeta$. This enforces $\omega_\zeta$ to be zero, i.e. $\vec{e}_\zeta$ must be perpendicular to $\vec{\omega} = \omega\vec{e}_z$. Introducing the Eulerian angles $\Phi, \Theta$ and $\Psi$ which connect the two coordinate systems we have the result $\Theta = \frac{\pi}{2}$ and consequently

$$B_\xi = B_0 cos\Psi cos\Phi + B_z sin\Psi \tag{25}$$

$$B_\eta = -B_0 sin\Psi cos\Phi + B_z cos\Psi \tag{26}$$

$$B_\zeta = B_0 sin\Phi \tag{27}$$

The quantities A=$a + ib$, C=$c + id$ may now be written down immediately as

$$a = \frac{1}{2}(B_\eta - \omega_\eta - \Gamma B_\xi) \tag{28}$$

$$b = \frac{1}{2}(B_\xi - \omega_\xi + \Gamma B_\eta) \tag{29}$$

$$c = \frac{1}{2}(B_\eta - \omega_\eta + \Gamma B_\xi) \tag{30}$$

$$d = -\frac{1}{2}(B_\xi - \omega_\xi - \Gamma B_\eta) \tag{31}$$

Observing that equations (23) and (24) are 3 real equations for the three unknowns $\Phi, \Psi$ and x we eliminate $\Phi$ and $\Psi$ and end up with equation (13) of the text.
Finally, in the context of the discussion of stability of these fixed points we need the following formulas: Using (23) to express $c$ and $d$ in terms of $a, b$ and $x$ we solve the four equations (28)-(31) for $B_\xi, B_\eta, \omega_\xi, \omega_\eta$ in terms of $a, b$ and $x$ as well. Noting that (24) is equivalent to $B_\zeta + 2J\frac{1-x^2}{1+x^2} = 0$ and $\omega_\zeta = 0$ we end up with

$$\begin{array}{ll} B_\xi = -\frac{a}{\Gamma}(1+x^2) & \omega_\xi = -(\frac{a}{\Gamma} + b)(1+x^2) \\ B_\eta = -\frac{b}{\Gamma}(1-x^2) & \omega_\eta = (\frac{b}{\Gamma} - a)(1-x^2) \\ B_\zeta = -2J\frac{1-x^2}{1+x^2} & \omega_\zeta = 0 \end{array} \tag{32}$$

As obviously $\omega = \sqrt{\omega_\xi^2 + \omega_\eta^2 + \omega_\zeta^2}, B_z = \frac{B_\xi \omega_\xi + B_\eta \omega_\eta}{\omega}$ and $B_0 = \sqrt{(\vec{B}_0 + \vec{B}_z)^2 - B_z^2}$ we have 3 relations which can be interpreted as effecting a transition from the original parameters $B_0, B_z, \omega$ to a new set of parameters $a, b, x$ leaving $J$ and $\Gamma$ fixed, and vice versa.

# Figure captions

figure 1
Bifurcations in the J-$\omega$ plane with $B_0 = 1.2, B_z = 1.0, \Gamma = 0.1$
$H_1$ and $H_2$ are the Hopf bifurcation lines of the trivial fixed point. At E there are four imaginary eigenvalues. SN are saddle-node bifurcation lines of nontrivial fixed points, inside the triangle A,$C_1, C_2$ there exist six nontrivial fixed points, outside only two. $C_1$ and $C_2$ are cusp points. $H_3$ and $H_4$ are Hopf bifurcation lines of the nontrivial fixed points. $F_1$ and $F_2$ have two vanishing eigenvalues.

figure 2
Bifurcations in the J-$\omega$ plane with $B_0 = 0.4, B_z = 1.0, \Gamma = 0.1$
$D_1$ and $D_2$ are bifurcations with two imaginary and one zero eigenvalue, Hopf and saddle node bifurcation lines are in tangential contact with each other.

figure 3
Bifurcations in the J-$\omega$-$B_0$ parameter space with $B_z = 1.0, \Gamma = 0.1$. Codimension 2 points of fig.(1),(2) appear as lines. Note that D is a closed curve. $F_1$ and $F_2$ terminate at $V_1$ and$V_2$

figure 4
Feigenbaum-scenario for $B_0 = 0.9, \omega = 0.9, B_z = 1.0, \Gamma = 0.1$. The Poincare plane is defined by $S_{x1} = 0$, $S_{y1}$ is plotted over the parameter J.

figure 5
Plot of the largest two Lyapunov exponents as functions of the parameter J for $B_0 = 0.9, \omega = 0.9, B_z = 1.0, \Gamma = 0.1$. The computation was carried out by a QR decomposition based method [14],[15].

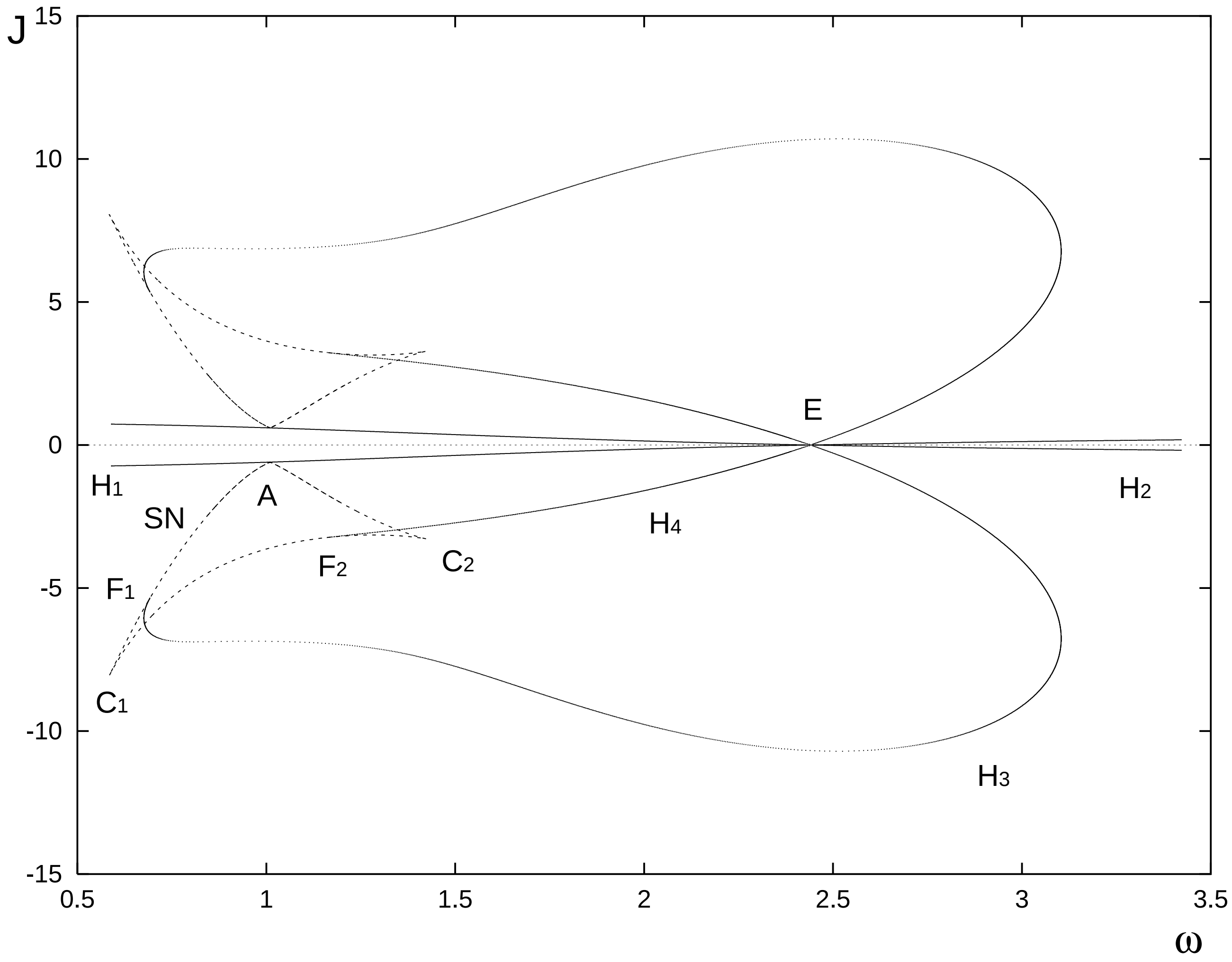

figure1

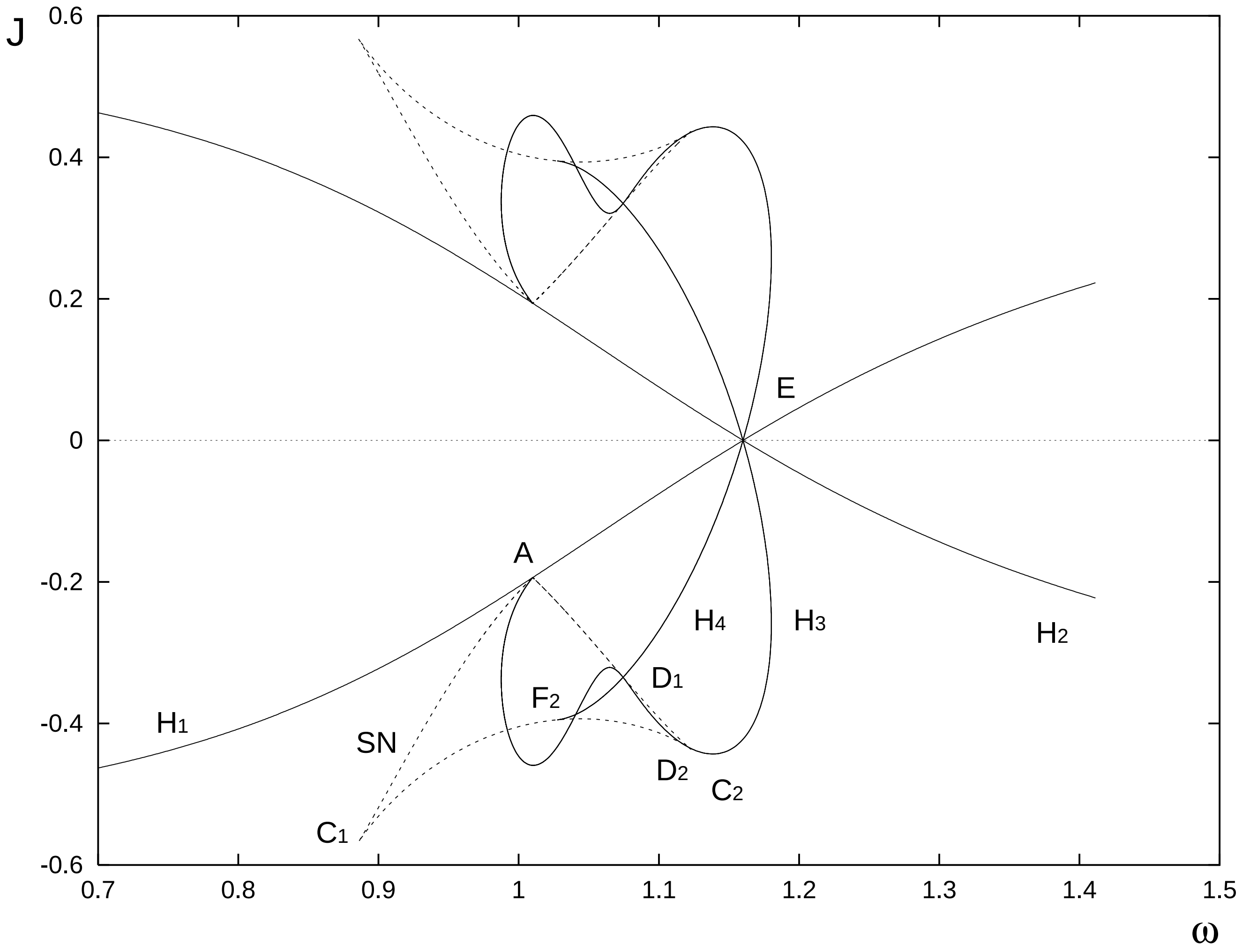

figure 2

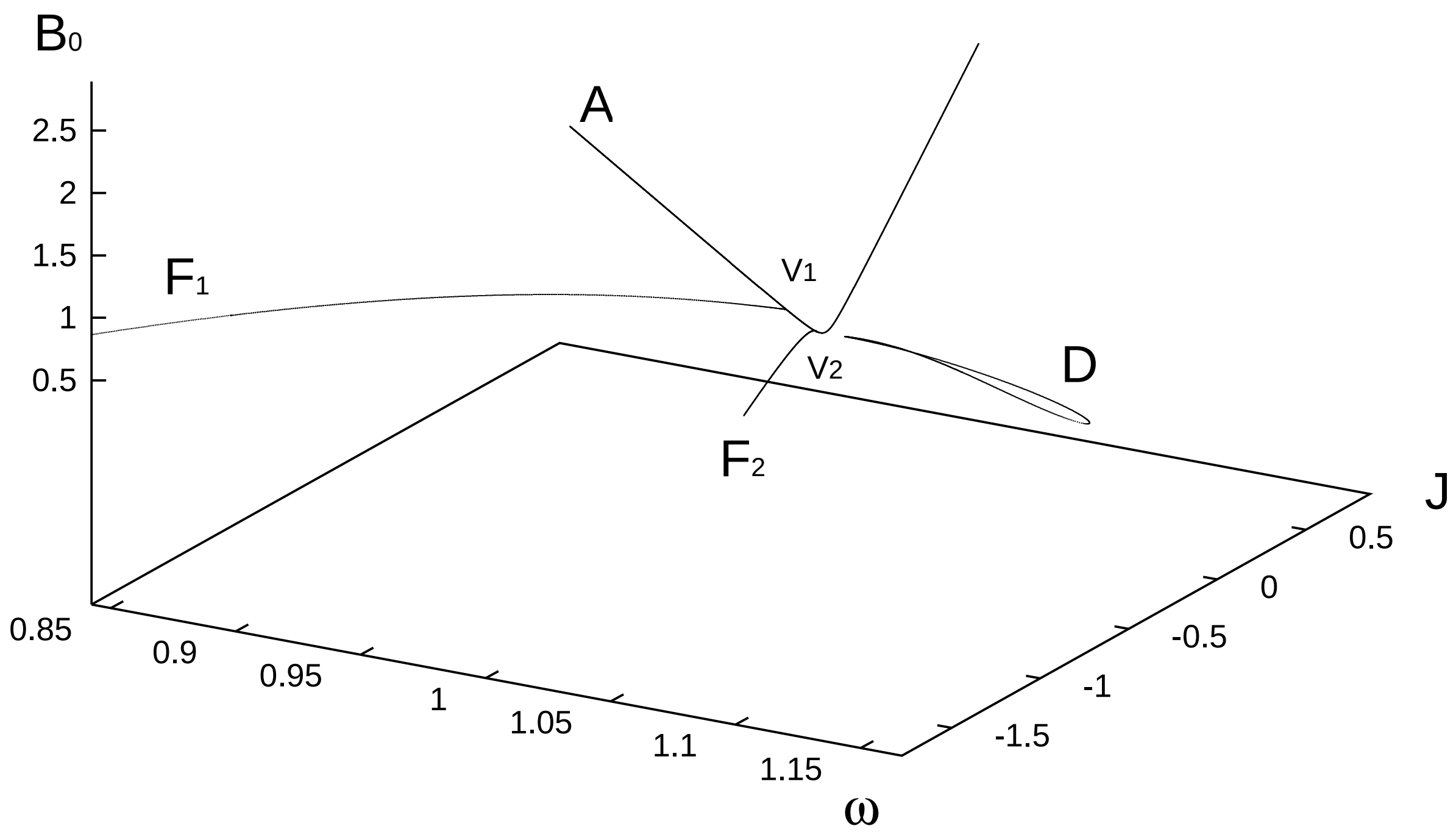

figure 3

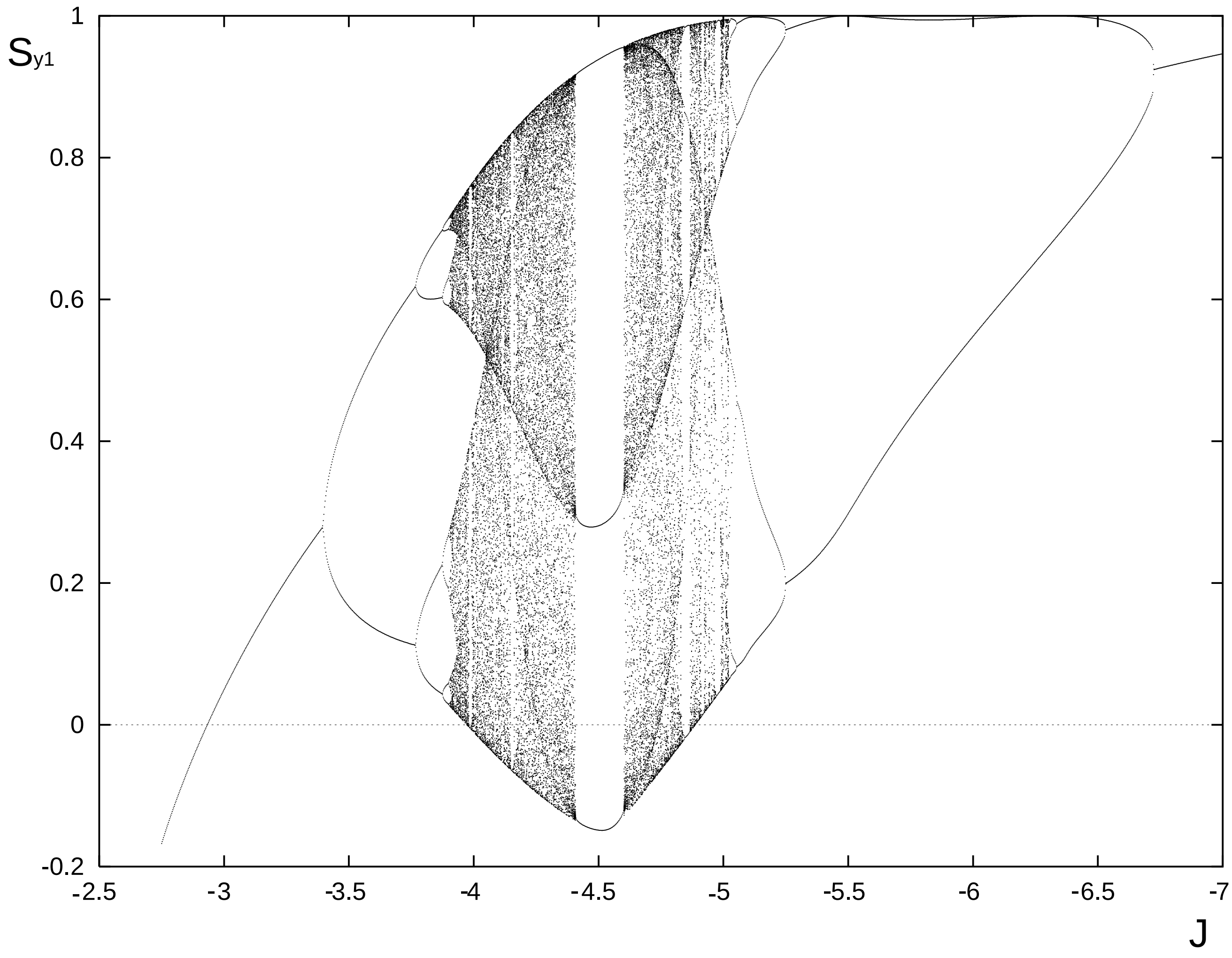

figure4

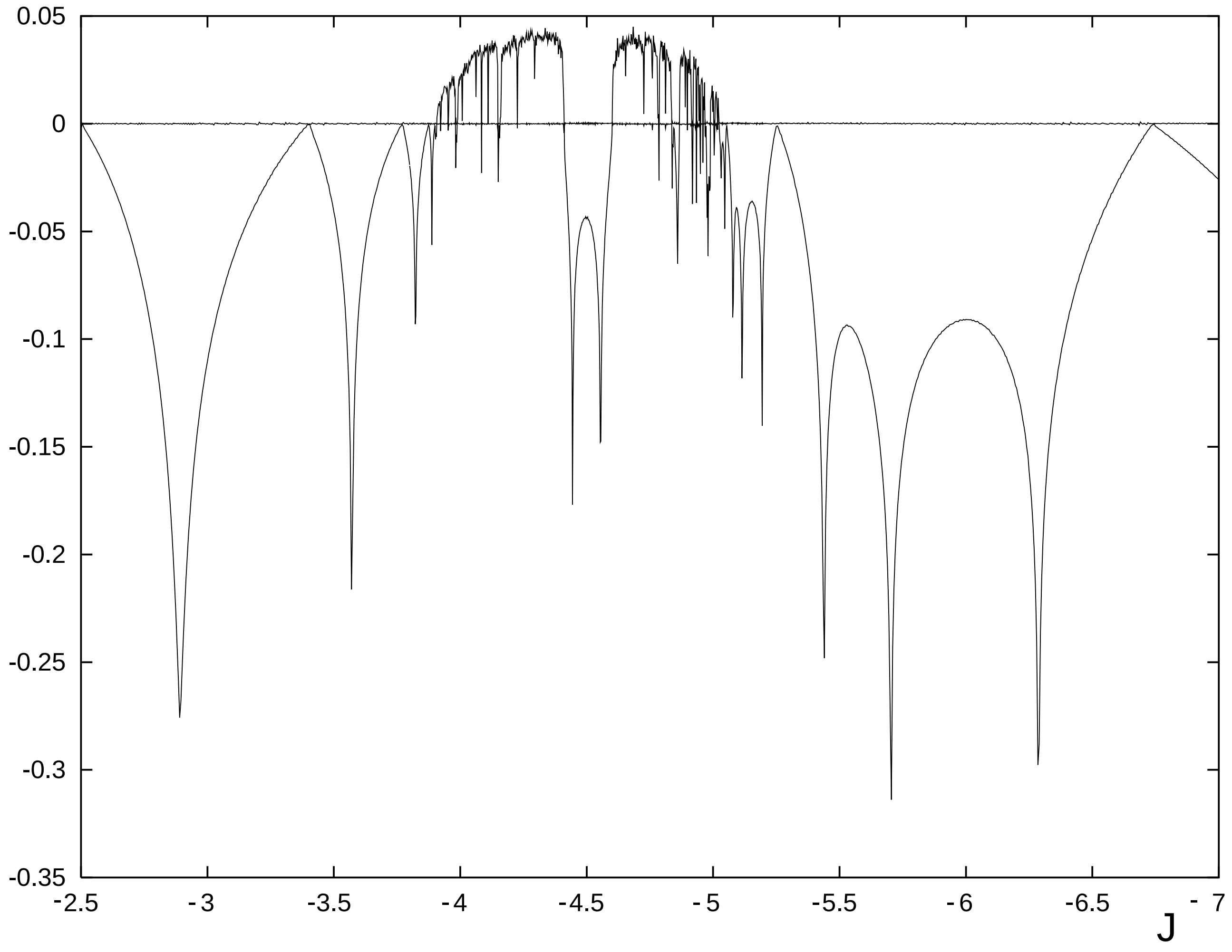

figure 5